\documentclass[a4paper,12pt]{article}
\usepackage[utf8]{inputenc}
\usepackage{cite}
\usepackage[margin=1.3 in]{geometry}
\usepackage{braket}
\usepackage{amsmath}
\usepackage{amsfonts}
\usepackage{hyperref}
\usepackage{graphicx}
\usepackage{comment}
\usepackage{xcolor}
\usepackage{amssymb}
\usepackage{authblk}
\usepackage{listings}
\usepackage{graphicx}  
\usepackage{subcaption}  
\usepackage{float}
\usepackage{xcolor}
\usepackage{listings}

\definecolor{keyword}{HTML}{0000FF}
\definecolor{string}{HTML}{008000}
\definecolor{comment}{HTML}{808080}
\definecolor{background}{HTML}{F5F5F5}

\title{\LARGE Quantum Zeno Dynamics of \\ Two Interacting Particles}

\author[1]{Varqa Abyaneh}
\author[2]{Parsa Ghorbani}

\affil[1]{\it\small Opetek, Level 37, 1 Canada Square, Canary Wharf, London, UK}
\affil[2]{\it\small Physics Department, Faculty of Science, Ferdowsi University of Mashhad, Iran}
\date{}

\begin{document}

\maketitle
\begin{abstract}
According to quantum Zeno dynamics (QZD), the evolution of a quantum system can be restricted to a subspace of its Hilbert space by frequent measurements.  
A crucial question in QZD of a particle's position is: how short the time interval between successive measurements should be, in order to confine the particle in  its initial spatial region? To address this question, we consider a toy model with two ions initially known to be, for simplicity, in a one-dimensional spatial region. By simulating the evolution of this two-body quantum system, we estimate the measurement frequency needed to keep the ions within their initial confined region at a desired confidence level. Two key parameters we employ in our calculations are the Zeno time and the leakage probability of the quantum system. The measurement frequencies are calculated and compared when ions are located initially at different spatial regions. For our simulation, we introduce the Python code {\tt 2IonQZD}. 
\end{abstract}

\newpage

\tableofcontents
\section{Introduction}
The precise manipulation and confinement of ions play a pivotal role in diverse fields of physics due to their wide range of applications \cite{romaszko_engineering_2020}. These include quantum computing \cite{bernardini2023quantum,cirac1995quantum}, quantum sensors \cite{baumgart2016ultrasensitive}, atomic clocks \cite{ludlow2015optical}, quantum simulators \cite{porras2004effective}, mass spectrometers \cite{dawson1976quadrupole}, and cold-atom experiments. For example, in quantum computing, trapped ions are among the most promising candidates for qubits (with companies such as IonQ \cite{schwaller_experimental_2022} and Quantinuum \cite{london_peptide_2023} having developed computers that utilize ion traps), alongside superconducting qubits \cite{Schwerdt:2023ion}. By isolating and manipulating individual ions, quantum operations can be performed with high precision, paving the way for fault-tolerant quantum computation \cite{quantum_computing_ion_benhelm_towards_2008,Gale_2020,wang2023faulttolerant}. This precision can be achieved using sophisticated quantum optimal control techniques \cite{werschnik2007quantum,Koch_2022,wang2022robust,bondar2022quantum,goerz2022quantum}.

Traditionally, ion confinement has been understood to require external forces or fields \cite{paul1990electromagnetic}, a notion accompanied by many challenges \cite{ion_trapping_wineland_experimental_1998}. Heating induced by field fluctuations poses a significant problem, compromising field fidelity and increasing the need for error correction \cite{bruzewicz_trapped-ion_2019}. Control is another issue, as dynamical fields can induce decoherence, which is detrimental to applications in which coherence is crucial \cite{trypogeorgos_synthetic_2018}. Scalability also poses problems as the number of ions increases \cite{zhang_versatile_2019}. Finally, achieving small confinement regions, such as those necessary for fusion reactions to occur, is impractical with traditional field-based methods due to the need for excessively strong fields.

Utilizing the Quantum Zeno Effect (QZE), a counterintuitive quantum phenomenon that allows the ``freezing'' of the state of a quantum system through frequent measurements \cite{QuantumZenoeffect,BALZER2002235,BALZER2000115}, it seems feasible to think of  trapping ions solely through measurements. 
However, trapping ions spatially by measurements requires a generalization of the QZE, known as Quantum Zeno Dynamics (QZD). 
In general, the QZD states that frequent measurements restrict the evolution of a quantum system not completely, but within a subspace of its Hilbert space \cite{QuantumZenosubspaces}. The experimental validation of QZD in finite-dimensional Hilbert space has already been demonstrated \cite{Sch_fer_2014}, but 
no experimental evidence exists for the infinite-dimensional Hilbert space case associated with position. 
While QZD in infinite-dimensional Hilbert space has been analytically shown for a single-particle system~\cite{Facchi_2008}, and Zeno dynamics in open systems and multi-particle settings has been extensively studied, particularly in the context of system-environment coupling~\cite{PhysRevA.90.012101,chaudhry2017quantumzenoantizenoeffects,chaudhry2016generalframeworkquantumzeno}, these works typically focus on dephasing control or decay suppression via environmental interactions. Moreover, the interplay between the Zeno and anti-Zeno effects has been explored in such contexts, showing that frequent measurements may either inhibit or accelerate evolution depending on system parameters. In contrast, our study presents a direct numerical simulation of QZD for a closed two-particle quantum system, demonstrating spatial confinement via frequent projective position measurements. To our knowledge, such simulations quantifying leakage probability and Zeno time for interacting particles in infinite-dimensional Hilbert space have not previously been reported.

In the pursuit of a QZD-based method to spatially confine multi-particle systems via frequent position measurements, we present Python code {\tt 2IonQZD}, which demonstrates the QZD for two interacting particles. The code can be extended to multi-particle systems with arbitrary interactions. We employ numerical methods, specifically the finite difference and Crank-Nicolson methods, to solve the partial differential equations in the  eigenstate equation and the system's time evolution. We then quantify the wavefunction's confinement in position space by defining the ``leakage probability'' and evaluating the ``Zeno time''. Based on these parameters, we estimate appropriate measurement frequencies for various confinement sizes to maintain the spatial confinement of the particles.

In Sec. \ref{secQZD}, we review the theoretical formulation of the QZE and QZD. In Sec. \ref{secnum}, we describe the numerical methods, specifically the finite difference and the Crank–Nicolson methods, used to solve the eigenstate and time-evolution problems for the two-particle quantum system. Sec. \ref{results} presents the numerical results and simulations generated by the code, including plots of the key parameters (the leakage probability and the Zeno time) for various scenarios. Finally, we conclude in \ref{conc}. A detailed description of the code {\tt 2IonQZD}, is provided in the Appendix \ref{code}.

\section{Quantum Zeno Dynamics} \label{secQZD}
The Quantum Zeno Effect (QZE) is a phenomenon in which a quantum system can be ``frozen'' in its initial state when subjected to sufficiently frequent measurements. Each time a measurement is performed on the system, the wavefunction collapses back into its initial state, effectively halting the system's evolution over time \cite{Misra:1976by}.
An extension of the QZE concept is the Quantum Zeno Dynamics (QZD), where frequent measurements constrain the system's evolution, without completely freezing it. Instead, the quantum system is constrained to evolve within a subspace of the Hilbert space. The Hilbert space can be either finite-- or infinite--dimensional. 
For instance, the position Hilbert space of a quantum particle system is infinite-dimensional.\cite{Facchi:2000bs,Facchi_2008}.
 As an application, QZD can be utilized to confine an ion within a desired spatial region. While experiments have demonstrated the QZE \cite{Balzer_2000,Balzer_2002,Toschek_2001,Wunderlich_2001}, 
no experiment has yet been proposed to spatially confine one or more particles using QZD. While several theoretical studies have explored QZD in single-particle systems \cite{Facchi_2008}, and multiparticle systems have been studied in the context of Zeno and anti-Zeno effects with system-environment coupling \cite{PhysRevA.90.012101,chaudhry2017quantumzenoantizenoeffects,chaudhry2016generalframeworkquantumzeno}, these works focus primarily on dephasing or decay control rather than direct spatial confinement. Our study numerically investigates the QZD of two interacting particles undergoing frequent position measurements in a closed system, which to the best of our knowledge remains unexplored.

Consider a quantum system initially in a pure state $\ket{\phi_0}$ governed by a non-perturbed Hamiltonian $H_0$ at time $t=0$. Adding an interaction to the system, the Hamiltonian changes to $H=H_0+H_\text{int}$. The system evolves over time with the new Hamiltonian. The probability that the system remains in its initial state after a single measurement at time $t$, is called the {\it survival probability} and is given by 
\begin{equation}
    p(t)=|\braket{\phi_0|e^{-iHt/\hbar}|\phi_0}|^2.
\end{equation}
Expanding the survival amplitude $A(t) = \langle \phi_0 | e^{-iHt/\hbar} | \phi_0 \rangle$ in a Taylor series for small $t$, we obtain,
\begin{equation}
A(t) = 1 - \frac{i t}{\hbar} \langle H \rangle - \frac{t^2}{2\hbar^2} \langle H^2 \rangle + \cdots,
\end{equation}
where the expectation values $\langle H \rangle = \langle \phi_0 | H | \phi_0 \rangle$ and $\langle H^2 \rangle = \langle \phi_0 | H^2 | \phi_0 \rangle$. The survival probability is then,
\begin{equation}
p(t) = |A(t)|^2 \approx 1 - \frac{t^2}{\tau_Z^2},
\end{equation}
which defines the Zeno time as,
\begin{equation}\label{zt}
\frac{\hbar^2}{\tau_Z^2} = \langle H^2 \rangle - \langle H \rangle^2.
\end{equation}
Now, if over a time interval $t$, we perform not a single measurement but $N$ measurements, the probability of finding the system in its initial state becomes
\begin{equation}
    p^{(N)}(t)\approx \left(  1-\frac{t^2}{N^2 \tau_Z^2}\right)^N.
\end{equation}
For a large number of measurements $N$, the survival probability effectively becomes an exponential function,
\begin{equation}\label{ztcon}
p^{(N)}(t)\approx \exp{\left(-\frac{t^2}{N^2\tau_Z^2}\right)}\to 1,~~~\text{if $\frac{t/N}{\tau_Z}\ll 1$},
\end{equation}
which implies that very frequent (almost continuous) measurements, ``freeze'' the quantum system in its initial state; a phenomenon called Quantum Zeno Effect (QZE).

The QZE can be generalized to quantum Zeno subspaces where incomplete measurements, represented by the operator $P$, project the quantum state of the system into a specific Hilbert subspace $\mathcal{H}_P$. 
If the initial quantum state is described by the density matrix $\rho_0$ within the Hilbert subspace $\mathcal{H}_P$, the time evolution of the system  results in the density matrix $\rho(\tau)=U(\tau)\rho_0 U^\dagger(\tau)$, where $U(\tau)=\exp{(-iH\tau)}$ is the time evolution operator, and $H$ is the Hamiltonian of the system. Now, if we measure the evolved density matrix $\rho(\tau)$ using the projection operator $P$, the survival probability i.e., the probability that the state is found again in the Hilbert subspace, is given by
\begin{equation}
    p(\tau)=\text{Tr}[\rho(\tau)P]=\text{Tr}[V(\tau)\rho_0 V^\dagger(\tau)].
\end{equation}
with $V(\tau)=Pe^{-iH\tau}P$. 
For multiple measurements, say $N$ times, every $\tau$ time, the survival probability after $t=N\tau$ is given by
\begin{equation}
    p^{(N)}(t)=\text{Tr}[V_N(t) \rho_0 V_N(t)]
\end{equation}
where $V_N(t)=(Pe^{-iHt/N}P)^N$.

In the limit $N \to \infty$, the repeated projections lead to an effective evolution given by
\begin{equation}
V_N(t) = \left(P e^{-i H t/N} P\right)^N \longrightarrow U_Z(t) = e^{-i H_Z t},
\end{equation}
where $H_Z = P H P$ is the so-called Zeno Hamiltonian. While $U_Z(t)$ is not unitary on the full Hilbert space, it is unitary within the Zeno subspace $\mathcal{H}_P = P \mathcal{H}$. In this limit, the survival probability becomes
\begin{equation}
\lim_{N \to \infty} p^{(N)}(t) = \text{Tr}\left[U_Z(t)\rho_0 U_Z^\dagger(t)\right] = \text{Tr}[\rho_0 P] = 1.
\end{equation}
The initial quantum state  we discussed above, can be any state within a finite- or infinite-dimensional Hilbert space. Whether the Hilbert space is finite- or infinite-dimensional depends on the specific system; for example, a two-level atom has a finite-dimensional Hilbert space, while the harmonic oscillator has an infinite-dimensional one. We are particularly interested in scenarios involving infinite-dimensional Hilbert spaces, such as the QZD for the position of particles.  Note that the quantum Zeno subspace can be proven analytically for infinite-dimensional Hilbert space and for a single particle \cite{Facchi:2000bs, Facchi_2001}. For a non-relativistic particle with mass $m$ in the potential $V(x)$, the Hamiltonian operator is given by $H=p^2/2m+V(x)$, where $p$ is the momentum operator.
A position state of the particle in infinite-dimensional Hilbert space is represented by $\ket{x}$ with $x\in \mathbb R$ in one-dimensional space. Now, measuring the particle's position in a compact subspace $\mathcal{R}$, is described by the projection operator, 
\begin{equation}
    P=\int_\mathcal{R} d x\ket{x}\bra{x}. 
\end{equation}
If the particle is found within the region $\mathcal{R}$, the projection operator $P$ preserves it; if it is at the boundary of or outside the region $\mathcal{R}$, it is projected to zero. As shown in Refs. \cite{Facchi:2000bs, Facchi_2001}, the evolution of the particle within the region $R$ is governed by the Zeno Hamiltonian $H_Z = P H P$, and the unitary evolution within the Zeno subspace is given by $U_Z(t) = \exp(-i H_Z t)$.
For very frequent measurements, the Zeno Hamiltonian operator $H_Z$ looks like 
\begin{equation}\label{qzd-1}
    H_Z= \frac{p^2}{2m}+V_\mathcal{R}(x)
\end{equation}
where the potential $V_\mathcal{R}(x)$ is defined as
\begin{equation}
V_\mathcal{R}(x) =
\begin{cases}
V(x), & x \in \mathcal{R}, \\
\infty, & \text{otherwise}.
\end{cases}
\end{equation}
Therefore, as seen from Eq. (\ref{qzd-1}), the evolution of a single particle, subjected to frequent position measurements, adds an additional potential term to the Schrödinger equation \cite{ QuantumZenodynamics}. This extra potential at the boundary of the compact region, acts as a ``hard wall'' or Dirichlet boundary, effectively confining the particle within the region.
Note that this effective infinite potential wall arises in the limit of infinitely frequent measurements and should be understood as an idealization. In practice, increasingly frequent projective measurements impose stronger boundary constraints, and in the limit, these mimic Dirichlet boundary conditions at the edge of the region $\mathcal R$, effectively creating a "hard wall" potential~\cite{Facchi_2001, Facchi_2008}.

\section{Time Evolution of the Wavefunction}\label{secnum}
We now describe the numerical model used to simulate QZD for a two-proton system. Our goal is to study the effect of frequent position measurements on the evolution of the wavefunction and estimate the conditions required for spatial confinement.

In general, the evolution of a non-relativistic two-body quantum system is governed by the time-dependent Schrödinger equation. For simplicity, we assume that each proton is allowed to move along only one spatial dimension, reducing the problem to a two-dimensional partial differential equation. Extending this approach to three spatial dimensions is straightforward, although the numerical calculations will be more computationally intensive. Because an analytical solution is typically not feasible, we employ the finite difference method \cite{Crank_Nicolson_1947,thomas1998numerical} to solve the differential equation to obtain the initial eigenstate. The time evolution of the wavefunction is obtained using the Crank-Nicolson method \cite{NumericalRecipes}.

Since we are ultimately interested in confining ions by means of the QZD to a region of physical space, we aim to demonstrate that QZD applies to a two-body quantum system by quantitatively estimating the appropriate measurement frequency to achieve this goal. In general, two particles may interact via an arbitrary potential, but for practical reasons we assume that two particles are actually two protons interacting though their  electrostatic potential. 

To demonstrate the viability of the QZD in a two-proton system with  measurements of their positions, it is necessary to solve the relevant Schrödinger equation. The Hamiltonian for this problem incorporates the kinetic energy of the protons and the potential energy resulting from their electrostatic repulsion.
The solution to the Schrödinger equation  for the two-body problem is studied analytically in \cite{DELARIPELLE19845}. However, we present a numerical solution to this problem,  as it allows us to calculate the key parameters such as the leakage probability and the Zeno time. 

The time-independent Schrödinger equation for two-body system is given by
\begin{equation}
H \psi(x_1,x_2) = E \psi (x_1,x_2),
\end{equation}
where
\begin{align}
H = -\frac{\hbar^2}{2m_1}\frac{\partial^2}{\partial x_1^2} -\frac{\hbar^2}{2m_2}\frac{\partial^2}{\partial x_2^2}  \notag + V(x_1,x_2).
\label{SE}
\end{align}
Here, $H$ is the Hamiltonian operator, representing the total energy of the system. $m_1$ and $m_2$ are the masses of the protons, and $\psi(x_1, x_2)$ is the wavefunction of the two-body system, depending on the positions $x_1$ and $x_2$ of the two protons. $E$ is the total energy eigenvalue of the system, and $V(x_1, x_2)$ is the potential energy function arising from the repulsive Coulomb interaction between the two protons,
\begin{equation}\label{vx1x2}
    V(x_1,x_2)= \frac{kq_1 q_2}{|x_1-x_2|},
\end{equation}
where $k= 8.99 \times 10^9$ N m$^2$/C$^2$.

Assuming that the protons  are initially confined in a one-dimensional region from $x_i=0$ to $x_i=L$, for $i=1,2$, the boundary conditions are given by,
\begin{equation}\label{bcs}
\psi(0,x_2)=\psi(x_1,0)=\psi(L,x_2)=\psi(x_1,L)=0.
\end{equation}
In the next subsections, we explain a method to solve the time-independent and time-dependent Schrödinger equations to evaluate key parameters of the leakage probability and the Zeno time.
\subsection{Eigenstates from Finite Difference Method}
Let us assume that the protons are in their ground state and confined in a spatial region. 
We choose the ground state as the initial state for simplicity. Since the ground state is the most localized and stable bound state of the system, it provides a starting point for evaluating how the wavefunction evolves. Moreover, the ground state has minimal initial energy and spatial spread, making it a natural reference for quantifying leakage and Zeno time. Nevertheless, our method can be applied to excited or arbitrary states.
To obtain the ground state wavefunction solution, we need to solve the time-independent Schrödinger eigenvalue problem $H\psi=E\psi$, which is a second-order partial differential equation (PDE). To achieve this, we discretize the continuous variables into a finite $N$ by $N$ grid using the finite difference method (FDM) \cite{thomas1998numerical}. 

Specifically, we assume the protons with positions $x_1$ and $x_2$ are confined to be in a one-dimensional potential well of length $d$, hence experiencing a hard wall potential with vanishing wavefunction at the boundaries (see Eq. (\ref{bcs})).

In this method, the continuous wavefunction $\psi(x_1,x_2)$ is replaced by a discrete $\psi(i,j)$ on the grid, where $(i,j)$ is a point on the grid corresponding to $(x_1,x_2)$. 
 We must ensure that the grid calculations are well-behaved.  Clearly, the potential in Eq. (\ref{vx1x2}) is not well-behaved for $x_1=x_2$. This can be rectified by regularizing $V(x_1,x_2)$,
\begin{equation}
\label{eq:Regularisation}
V(x_1,x_2) =  \frac{k q_1q_2}{\sqrt{(x_1-x_2)^2+\epsilon^2}},
\end{equation}
where $q_1$ and $q_2$ are the respective charges of each proton. The regularization parameter $\epsilon$, is set to $10^{-15}$ m. This is justified by noting that when the denominator of Eq. (\eqref{eq:Regularisation}) is below this number, we expect a strong nuclear force to overcome the Coulomb repulsion \cite{wong1998introductory}. 
Now the eigenvalue problem reduces to a set of algebraic equations on the grid, 
\begin{equation}
\begin{split}
&-\frac{\hbar^2}{2}\left[ \frac{\psi(i+1,j)+\psi(i-1,j)-2\psi(i,j)}{m_1\Delta x^2} \right. \notag  
           \left. + \frac{\psi(i,j+1)+\psi(i,j-1)-2\psi(i,j)}{m_2\Delta x^2}\right]  \notag \\
&+ k \frac{q_1q_2}{\sqrt{(i-j)^2\Delta x^2 +\epsilon^2}}\psi(i,j) \notag =E\psi(i,j),\hspace{1cm}i,j=1,2,...,N-1
\end{split}
\end{equation}
where $\Delta x = L / N$ is the distance between two successive points on the grid. 

 The boundary conditions translated from the continuous form in Eq. (\ref{bcs}) into the grid form become
 \begin{equation}
     \psi(0,i)= \psi(i,0)= \psi(N,i)= \psi(N,i)=0
 \end{equation}
 for $i=0,1,..,N$.
 We have one equation for each ($i,j$) pair, and the eigenvectors and associated eigenvalues can be obtained numerically. 

\subsection{Crank--Nicolson Time Evolution}

To grasp the required frequency of measurements to observe the QZD effect, it is crucial to study the time evolution of the two-proton quantum system. 
The idea is to perform a measurement to check whether both protons are confined within the desired spatial region at each time interval $\tau$. 
Therefore, we need to solve the time-dependent Schrödinger equation, $i\hbar\partial\psi/\partial t =H\psi$.  In order to do that, we exploit the Crank--Nicolson method, which is a stable numerical approach based on the finite difference method.
The Crank--Nicolson scheme for the time-evolving system is given by \cite{NumericalRecipes},
\begin{equation}\label{C-N}
\left( {I}-\frac{i\Delta t}{2\hbar} {H} \right) \psi(i,j,k+1) 
   = \left( {I}+\frac{i\Delta t}{2\hbar}{H} \right)\psi(i,j,k),
\end{equation} 
where ${I}$ is the identity matrix, ${H}$ is the Hamiltonian matrix, 
$\psi(i,j,k)$ is the state with $i,j$ representing the spatial grid point of each proton and $k$ representing the time step, and $\Delta t$ is the size of the time step (to be distinguished from the time interval between measurements $\tau$). 

In our simulations, the confinement region $R$ is defined as a square of side length $d$. The value of $d$ is a tunable parameter that determines the initial confinement area, and the time evolution of the wavefunction is used to assess how quickly it spreads outside this region.

\subsection{Leakage Function} \label{sec:Leakage function}
Instead of the survival probability discussed in Sec. \ref{secQZD}, we introduce the related quantity \textit{leakage function}. The leakage function is a related concept that serves as a measure of how much of the wavefunction has extended beyond a predefined boundary during the time between two successive measurements, $\tau$. This is useful for understanding the probability that the quantum system is confined within a certain region, thereby calibrating a suitable value for the frequency of confinement measurements, $f_{\text{QZD}}=1/\tau_{\text{QZD}}$, to observe QZD. 

In the continuous form, the leakage $L(\psi_\mathcal{R},\tau_{\text{QZD}})$ is defined as
\begin{equation}
\label{eq:leakage_function}
L(\psi_{\mathcal{R}},t_{QZD})= \int_{\text{outside}} |\psi_{\mathcal{R}}(x_1, x_2, t_{QZD})|^2 \, dx_1 \, dx_2.
\end{equation}
Here, $\psi_{\mathcal{R}}(x_1, x_2, \tau_{\text{QZD}})$ represents the wavefunction (previously measured and confined within the region defined by $\mathcal{R}$) of a proton at position $x_1$ and the other proton at position $x_2$ after the time $\tau_{\text{QZD}}$ has passed, when another measurement takes place to determine the positions of the protons. The integral is calculated over and outside the region $\mathcal{R}$.  The region $R$ is defined by the confinement length $d$, and the leakage is calculated by summing the probability density outside this region in the extended grid.

In the discrete form, the leakage $L(\psi_\mathcal{R},\tau_{\text{QZD}})$ can be expressed as
\begin{equation}
L(k) = \sum_{\text{outside}} |\psi_\mathcal{R}(i, j,k)|^2.
\end{equation}
In this equation, $\psi(i, j, k)$ is the value of the wavefunction at the grid point $(i, j,k)$, where $k$ is the time grid point associated with $\tau_{\text{QZD}}$. The sum is taken over grid points that lie outside the initial grid (As in the continuous case, the integral is over the region outside that defined by $\mathcal{R}$).
What we demonstrate numerically in the next section is that the leakage function vanishes when the measurement time interval tends to zero, 
\begin{equation}
\lim_{{\tau_{\text{QZD}} \to 0}} L(\psi_\mathcal{R}, \tau_{\text{QZD}}) = 0.
\end{equation}
Having $L(\psi_{d},\tau_{\text{QZD}}) = 0$ is equivalent to the frequency of measurements $f_{\text{QZD}} \to \infty$. Note that this is equivalent to having an infinite potential wall at the boundary of the region $\mathcal{R}$ in Eq. (\ref{qzd-1}).

\section{QZD of Two Protons: Numerical Results}\label{results}
\begin{figure}
    \centering
    \includegraphics[width=0.7\linewidth]{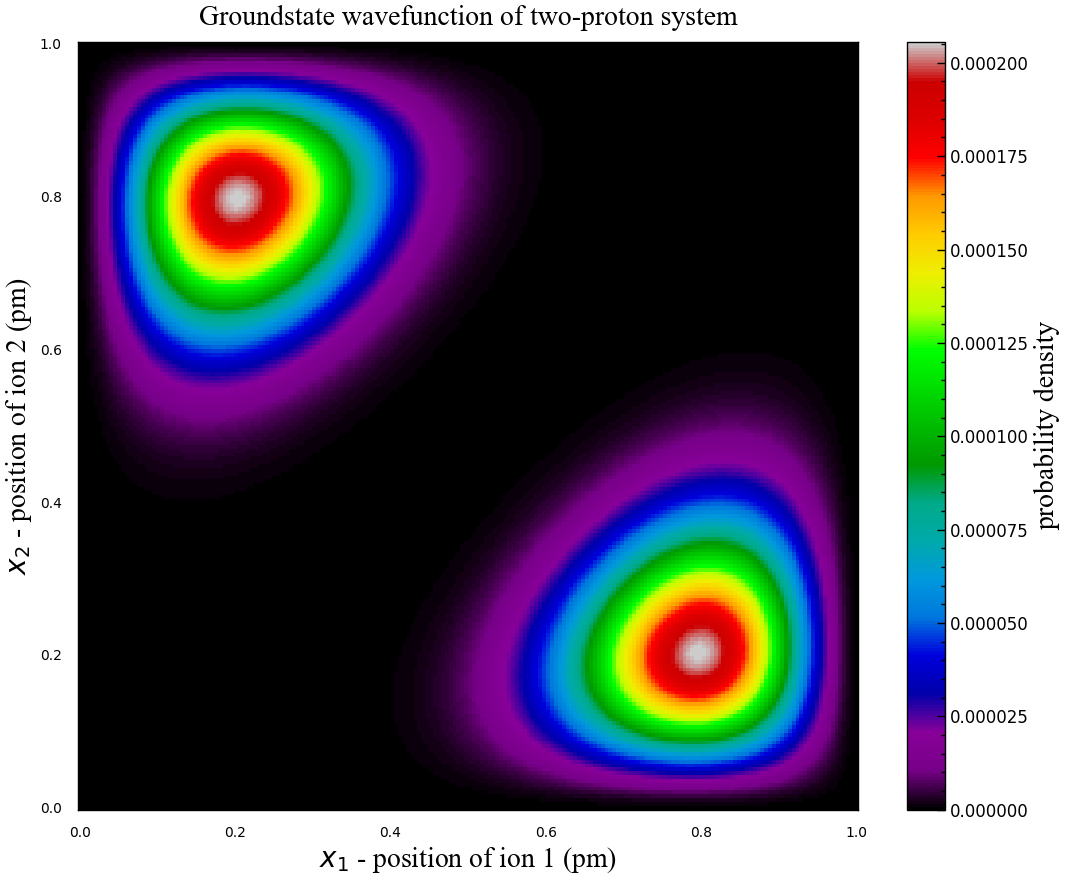}
    \caption{The ground  state probability density associated with the wavefunction solution to the Schrödinger equation for the two-proton system with electrostatic interaction in pico meter scale.}
    \label{fig1}
\end{figure}

As pointed out in the previous section, we assume that the ions are protons with their known charge and mass. 
The goal here is to demonstrate the effect of frequent measurements on the evolution of the wavefunction. According to QZD, if we choose the frequency of the measurements to be large enough, the wavefunction will remain confined within the Hilbert subspace defined by the measurement process. To achieve this aim, we calculate the leakage probability of the wavefunction as a function of time for different confinement sizes. 
Although the Coulomb repulsion pushes the protons apart, the confinement size $d$ is chosen such that the ground state remains well localized within the region $\mathcal R$, ensuring the initial leakage is negligible.
We also numerically compute the Zeno time defined in Eq. (\ref{zt}).  
The numerical calculations are performed by our code  in Python called {\tt 2IonQZD}. A detailed description of the code is provided in Appendix \ref{code}.
Running the code, the first result we obtain is the solution to the time-independent Schrödinger equation, which is a two-body eigenstate problem. In the code, we set the distance between two protons (confinement size) to {\tt d=1e-12} m, and specify the ground state solution by setting {\tt selected\_eigenstate = 0}. The result in Figure \ref{fig1}, shows the ground state probability density in color legend, in terms of the protons' positions. It is evident that the probability density is symmetric in the protons' positions and as expected it is close to zero on the line $x_1=x_2$ where the Coulomb repulsion is maximum.

Now, taking the ground state wavefunction as a starting point, the code computes the time evolution of the wavefunction elaborated in Sec. \ref{code}. 
The key parameter in our computation is the leakage probability which quantifies the extent to which the wavefunction penetrates outside the confinement region over time.  By evaluating the leakage probability as a function of time for different confinement region sizes $d$, we  estimate the appropriate measurement time intervals required to ensure the protons remain confined in various scenarios.  

Note that in our simulations, we do not explicitly perform wavefunction collapse after each time step, but instead compute the time evolution under the full Hamiltonian and evaluate the leakage probability at regular intervals. This models how the system would evolve between successive projective position measurements. The goal is to estimate how small the measurement time interval $\tau$ must be so that the leakage remains below a desired threshold---effectively mimicking the role of frequent measurements in Quantum Zeno Dynamics. 

\begin{figure}[htbp]
    \centering
    \begin{tabular}{ccc}
        \begin{subfigure}[b]{0.3\textwidth}
            \includegraphics[width=\textwidth]{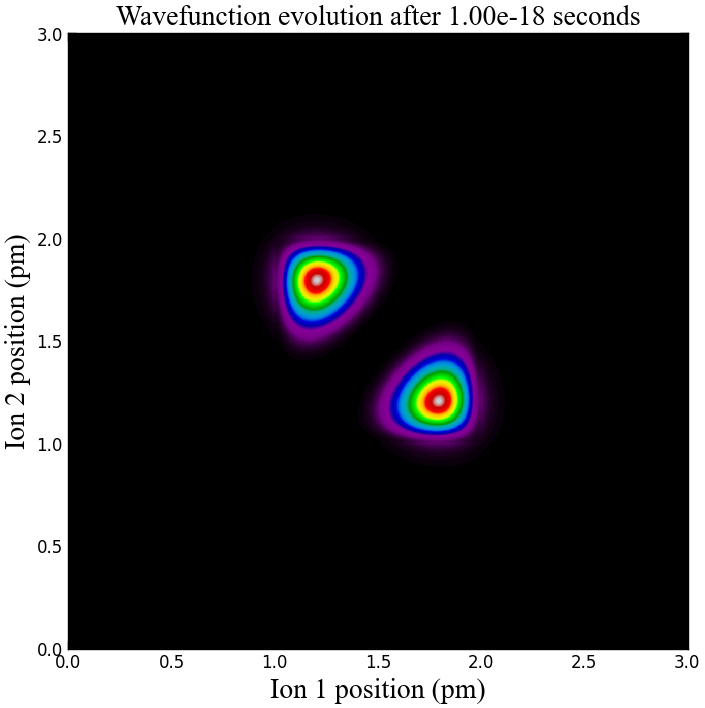}
            \caption{\small  $t=10^{-18}$ s}
        \end{subfigure} &
        \begin{subfigure}[b]{0.3\textwidth}
            \includegraphics[width=\textwidth]{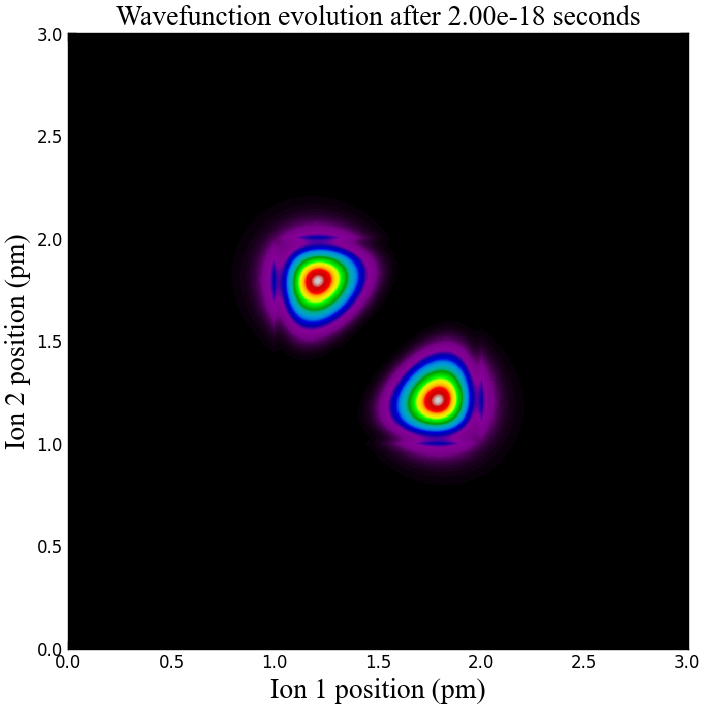}
            \caption{\small  $t=2\times 10^{-18}$ s}
        \end{subfigure} &
        \begin{subfigure}[b]{0.3\textwidth}
            \includegraphics[width=\textwidth]{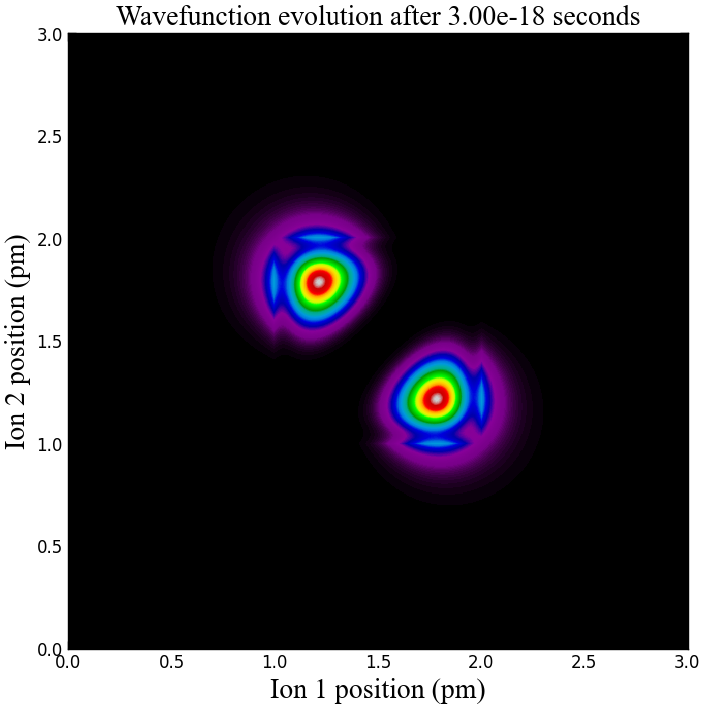}
            \caption{\small  $t=3\times 10^{-18}$ s}
        \end{subfigure} \\
        
        \begin{subfigure}[b]{0.3\textwidth}
            \includegraphics[width=\textwidth]{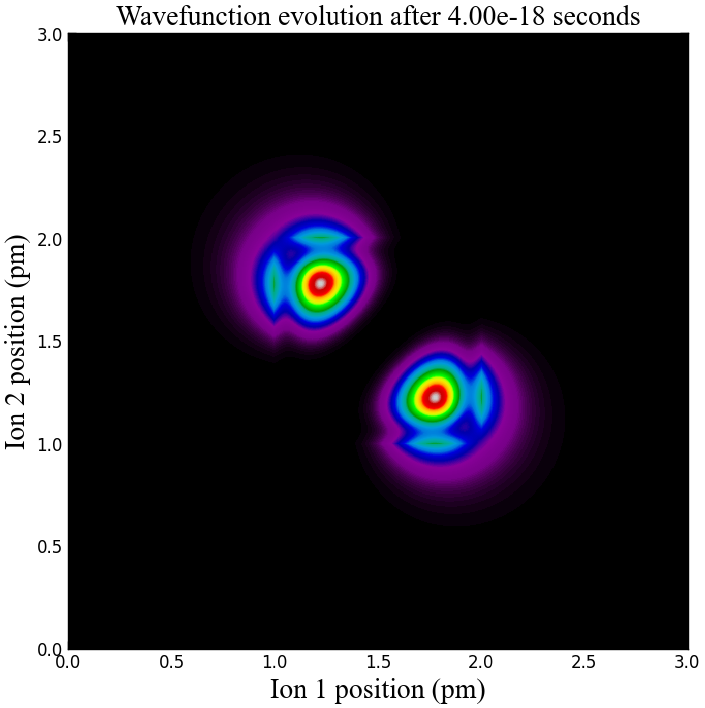}
            \caption{\small  $t=4 \times 10^{-18}$ s}
        \end{subfigure} &
        \begin{subfigure}[b]{0.3\textwidth}
            \includegraphics[width=\textwidth]{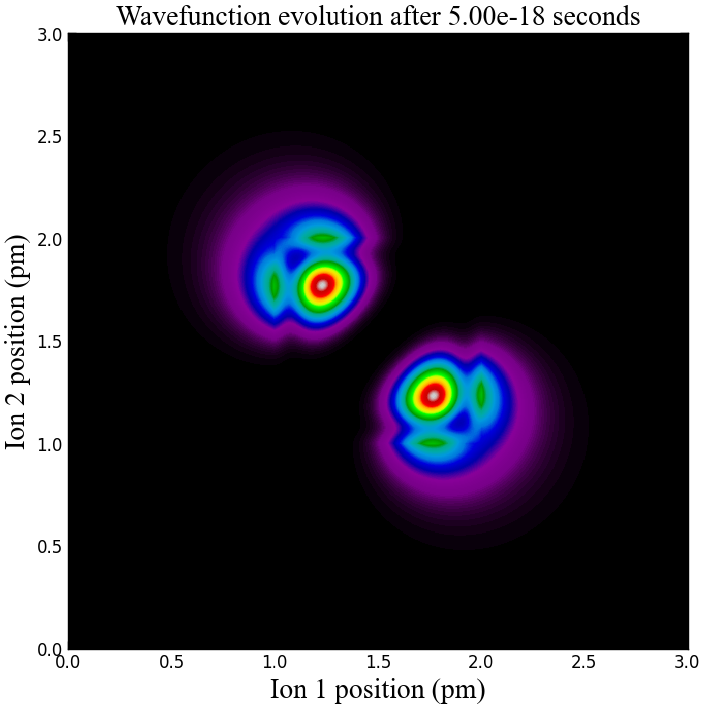}
            \caption{\small  $t=5 \times 10^{-18}$ s}
        \end{subfigure} &
        \begin{subfigure}[b]{0.3\textwidth}
            \includegraphics[width=\textwidth]{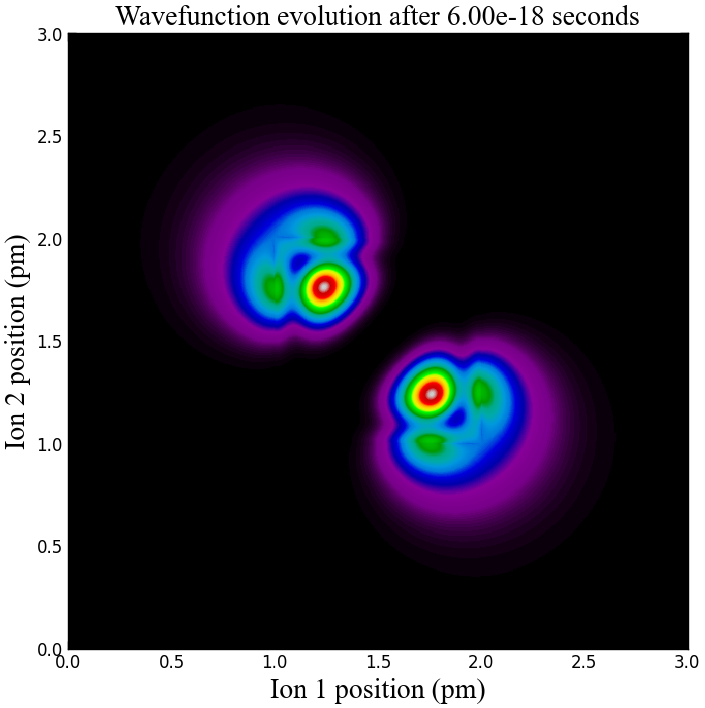}
            \caption{\small  $t=6\times 10^{-18}$ s}
        \end{subfigure} \\
        
        \begin{subfigure}[b]{0.3\textwidth}
            \includegraphics[width=\textwidth]{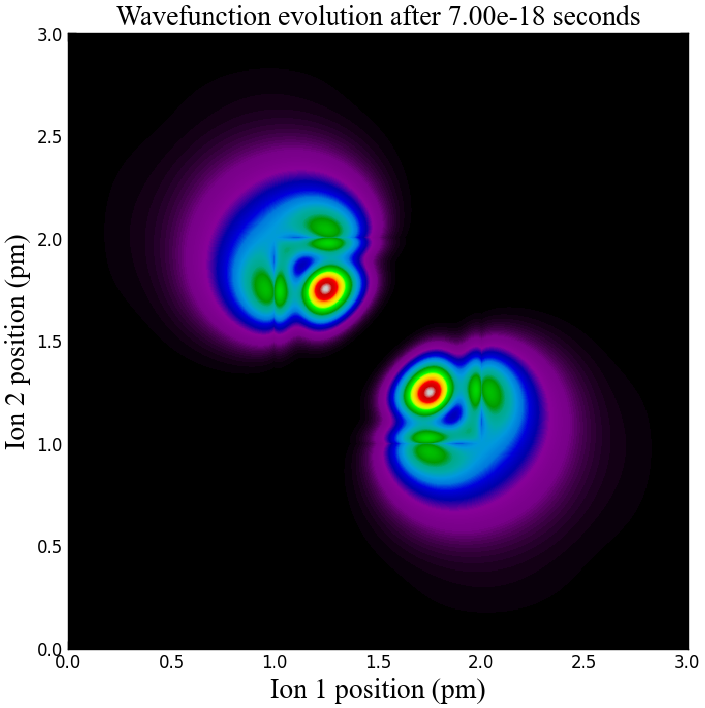}
            \caption{\small  $t=7\times  10^{-18}$ s}
        \end{subfigure} &
        \begin{subfigure}[b]{0.3\textwidth}
            \includegraphics[width=\textwidth]{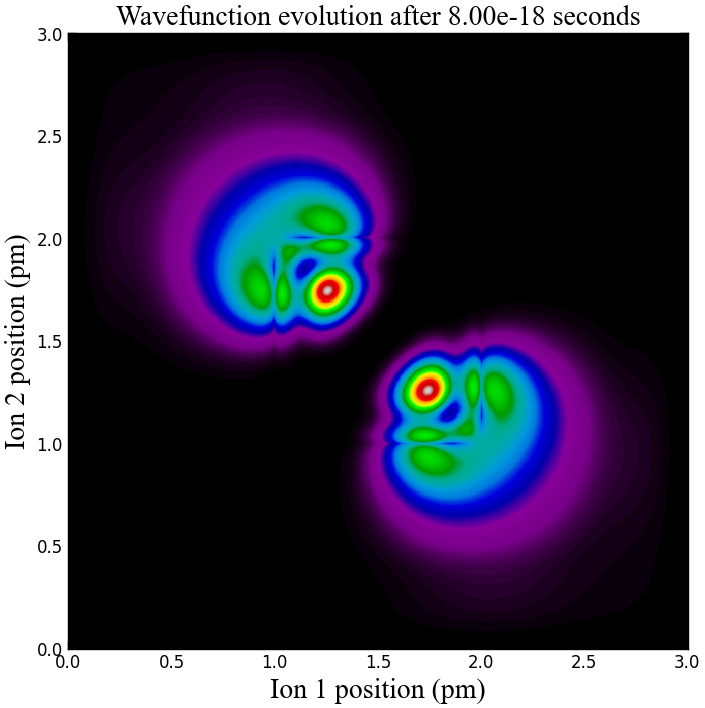}
            \caption{\small  $t=8\times 10^{-18}$ s}
        \end{subfigure} &
        \begin{subfigure}[b]{0.3\textwidth}
            \includegraphics[width=\textwidth]{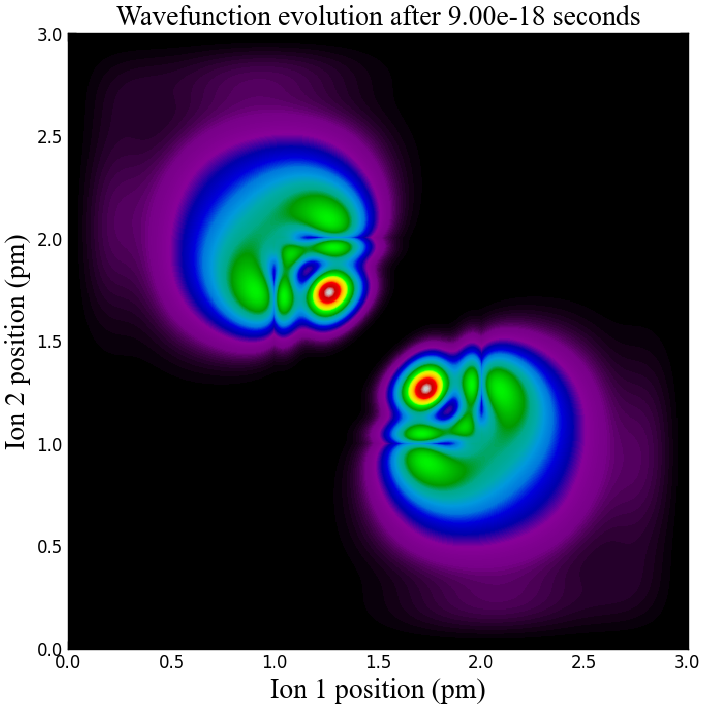}
            \caption{\small  $t=9\times 10^{-18}$ s}
        \end{subfigure}
    \end{tabular}
    \caption{The probability density against the ions' positions is shown for nine time steps $10^{-18}$ s. The wavefunction is mostly collapsed after $10^{-17}$ s.}
    \label{fig2}
\end{figure}

First, for the confinement region $d=10^{-12}$ m, we fix the time step $\delta t=10^{-18}$ and let the wavefunction evolve. For nine time steps, we show the probability density versus the protons' positions in Figure \ref{fig2}. As seen from this figure, the wavefunction has spread beyond its initial confinement region after $9 \times 10^{-18} \simeq 10^{-17}$ s, which means that the probability of both protons being confined inside the region defined by $d=10^{-12}$ m is small.  This figure shows qualitatively that selecting the measurement time interval as large as $10^{-17}$ s is not practical for confining the protons in the region $d=10^{-12}$ m.

We then examine the impact of smaller time steps on the evolution of the wavefunction. Let us choose $\delta t=10^{-21}$ s, i.e., a time step two orders of magnitude smaller than the previous choice. We calculate again the evolution of the wavefunction after a given number of time steps. In Figure \ref{fig4}, the wavefunction evolution after nine time steps is shown as the probability density versus the protons' positions. It is easily seen  that the wavefunction has not evolved much and both protons are still likely to be in the confinement region. Again, this figure qualitatively demonstrates that selecting a measurement frequency as small as $9 \times 10^{21} \simeq 10^{-20}$ s, results in both ions remaining confined in the region $d=10^{-12}$ m. 

\begin{figure}[htbp]
    \centering
    \begin{tabular}{ccc}
        \begin{subfigure}[b]{0.3\textwidth}
            \includegraphics[width=\textwidth]{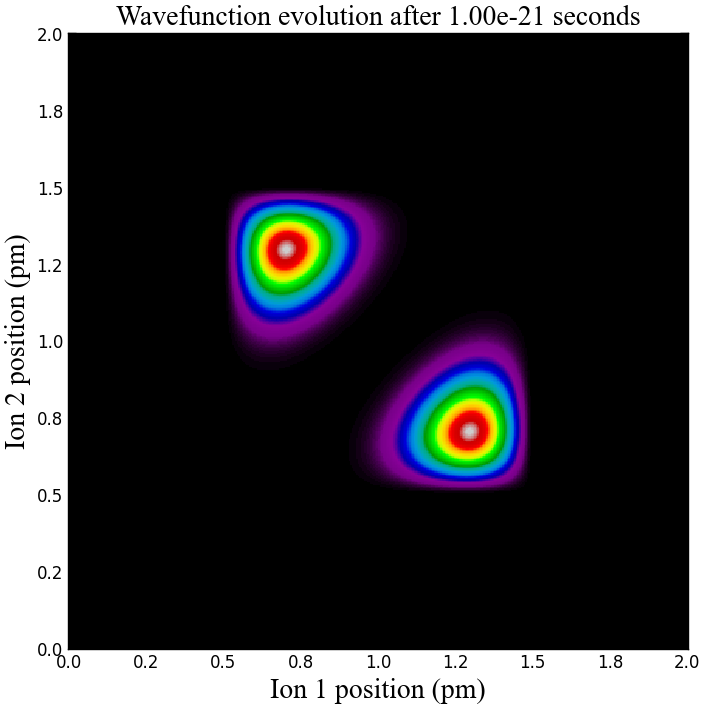}
            \caption{\tiny  $t=10^{-21}$ s}
        \end{subfigure} &
        \begin{subfigure}[b]{0.3\textwidth}
            \includegraphics[width=\textwidth]{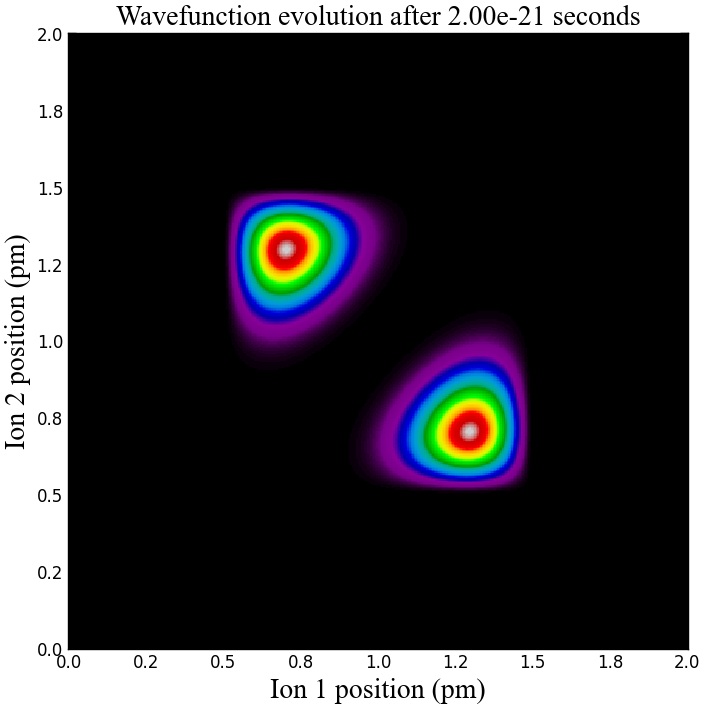}
            \caption{\tiny  $t=2\times 10^{-21}$ s}
        \end{subfigure} &
        \begin{subfigure}[b]{0.3\textwidth}
            \includegraphics[width=\textwidth]{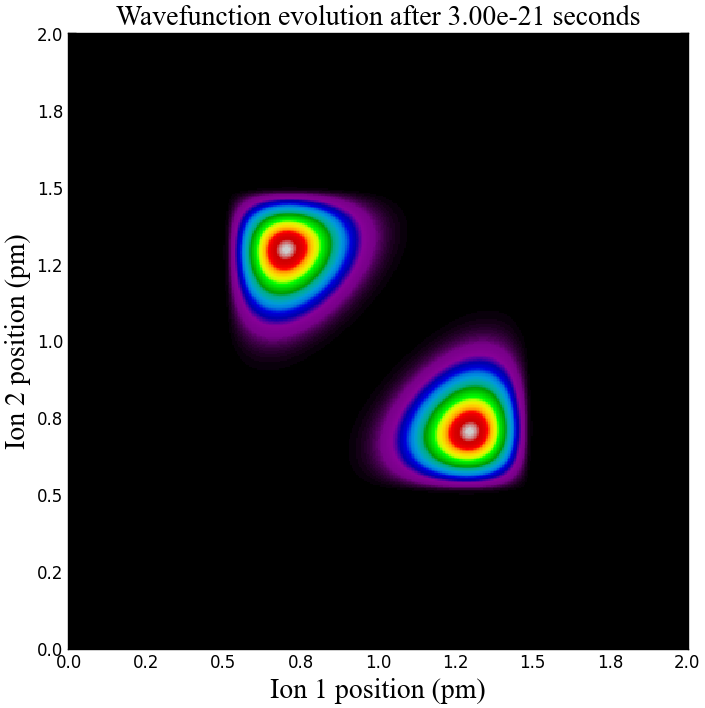}
            \caption{\tiny  $t=3\times 10^{-21}$ s}
        \end{subfigure} \\
        
        \begin{subfigure}[b]{0.3\textwidth}
            \includegraphics[width=\textwidth]{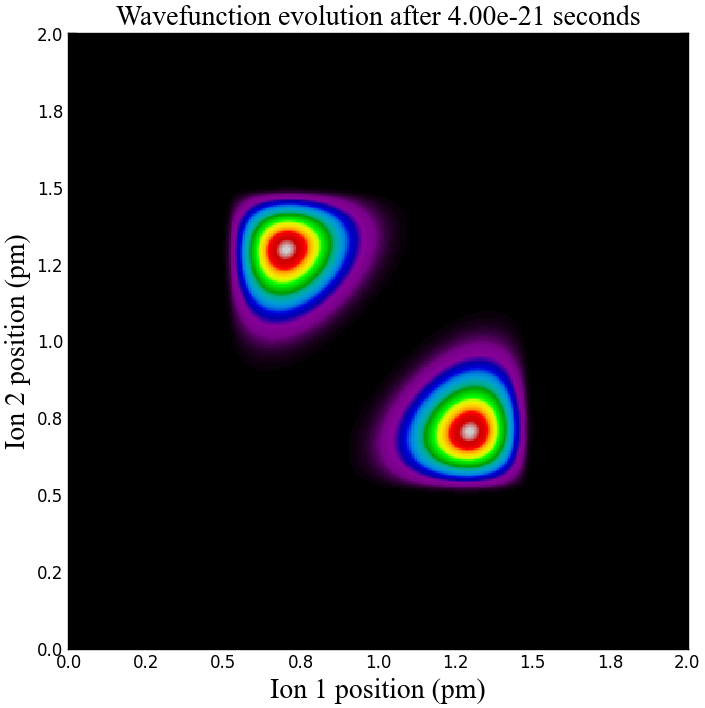}
            \caption{\tiny  $t=4 \times 10^{-21}$ s}
        \end{subfigure} &
        \begin{subfigure}[b]{0.3\textwidth}
            \includegraphics[width=\textwidth]{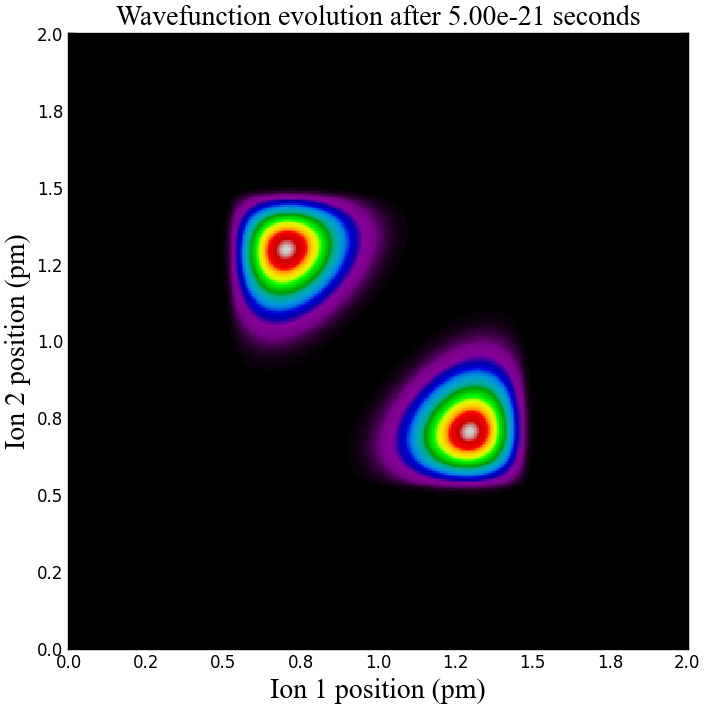}
            \caption{\tiny  $t=5 \times 10^{-21}$ s}
        \end{subfigure} &
        \begin{subfigure}[b]{0.3\textwidth}
            \includegraphics[width=\textwidth]{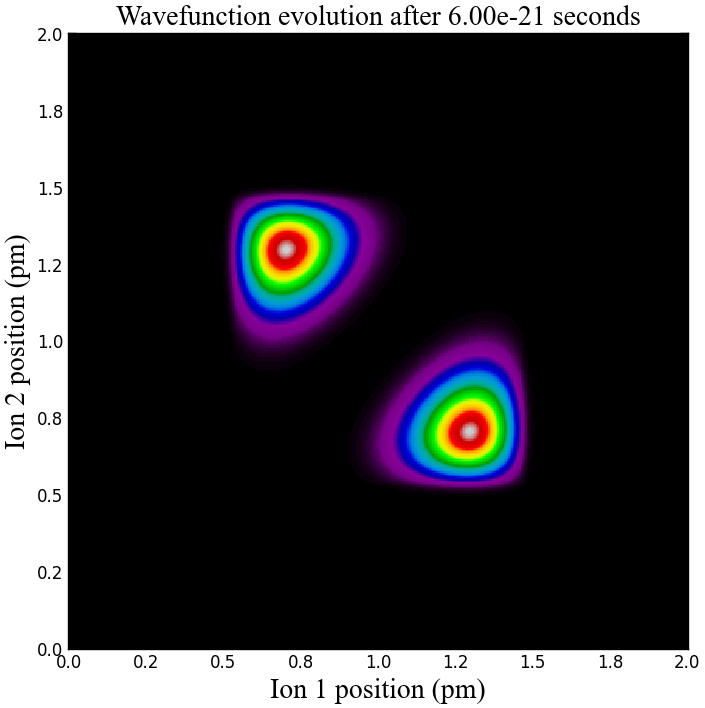}
            \caption{\tiny  $t=6\times 10^{-21}$ s}
        \end{subfigure} \\
        
        \begin{subfigure}[b]{0.3\textwidth}
            \includegraphics[width=\textwidth]{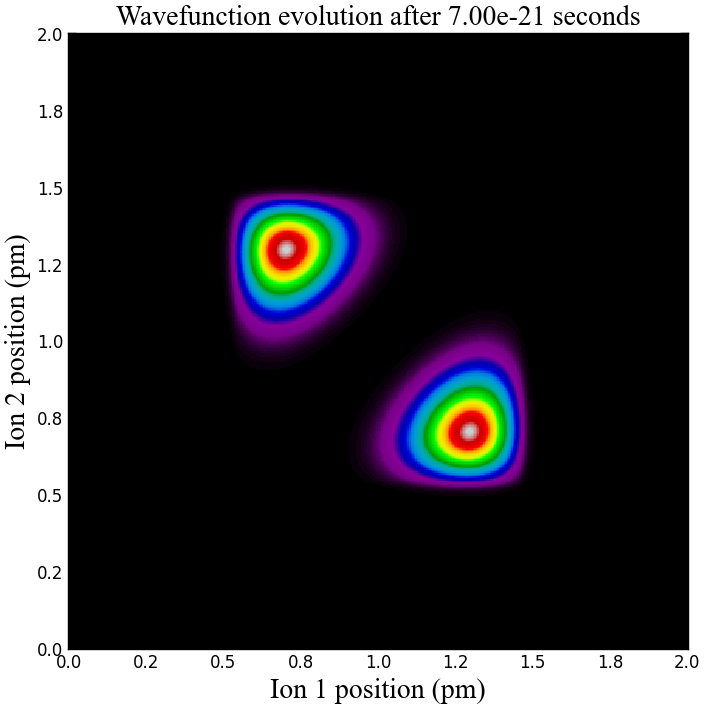}
            \caption{\tiny  $t=7\times  10^{-21}$ s}
        \end{subfigure} &
        \begin{subfigure}[b]{0.3\textwidth}
            \includegraphics[width=\textwidth]{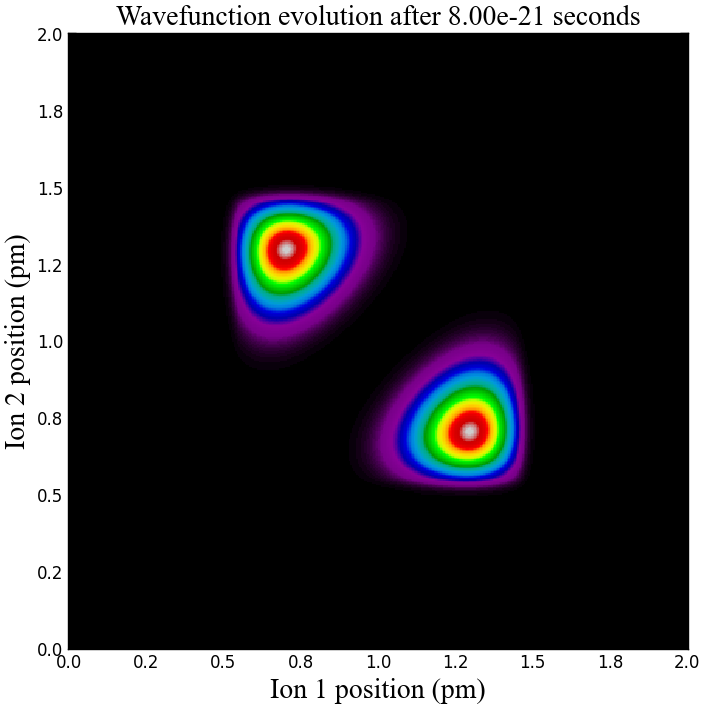}
            \caption{\tiny  $t=8\times 10^{-21}$ s}
        \end{subfigure} &
        \begin{subfigure}[b]{0.3\textwidth}
            \includegraphics[width=\textwidth]{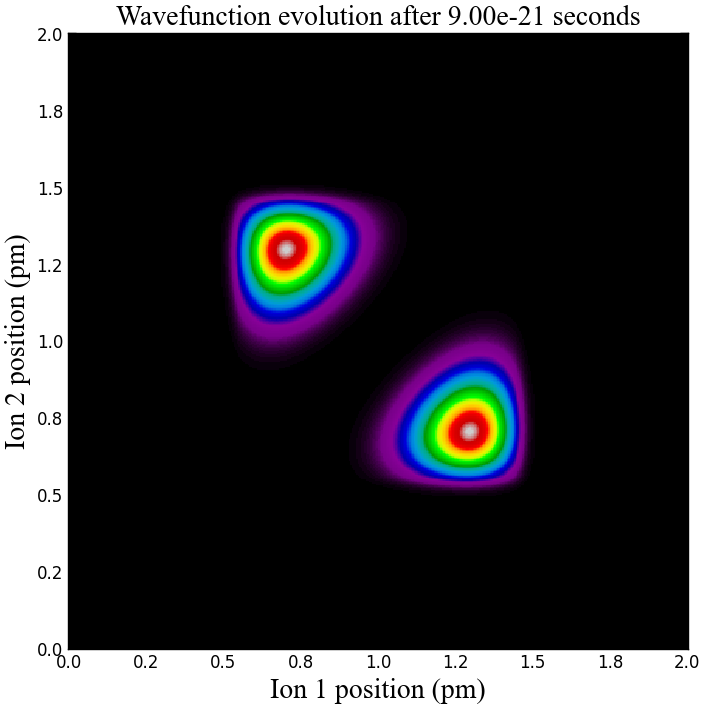}
            \caption{\tiny  $t=9\times 10^{-21}$ s}
        \end{subfigure}
    \end{tabular}
    \caption{The probability density versus the protons' positions is shown for nine time steps, $10^{-21}$ s. The ions are  observed with high probability inside the confinement region.}
    \label{fig3}
\end{figure}

Figures \ref{fig2} and \ref{fig3} present only the qualitative behavior of the wavefunction over time. 
In order to calibrate the measurement time interval, we need to calculate the leakage probability for different confinement sizes. The leakage probability for a single measurement is defined in Eq. (\ref{eq:leakage_function}). Running the code for different values of the confinement sizes $d=10^{-10}$ m, $d=10^{-11}$ m, and $d=10^{-12}$ m, the leakage probability is obtained for each scenario. The result is depicted in Figure \ref{fig4}. As expected, the leakage probability always increases over time independent of the confinement size. It is evident from this figure that when the protons are closer, due to the stronger repulsive force, the leakage probability approaches one more quickly. For  $d=10^{-12}$ m, the probability that at least one of the ions is outside the confinement region at the time $2\times 10^{-18}$s, is close to one. This time is $6\times 10^{-17}$ s and $6\times 10^{-15}$ s for $d=10^{-11}$ m and $d=10^{-10}$ m, respectively.

\begin{figure}[t]
    \centering
    \begin{tabular}{ccc}
        \begin{subfigure}[b]{0.3\textwidth}
            \includegraphics[width=\textwidth]{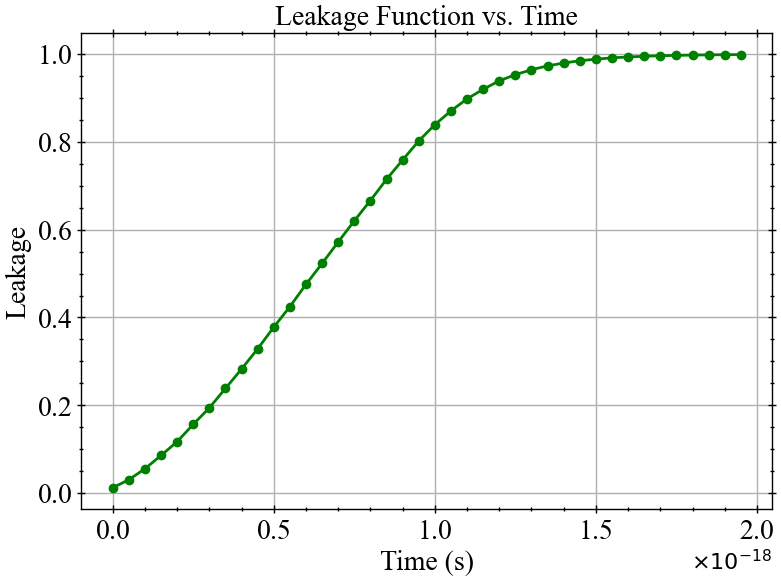}
            \caption{\small $d=10^{-12}$ m}
        \end{subfigure} &
        \begin{subfigure}[b]{0.3\textwidth}
            \includegraphics[width=\textwidth]{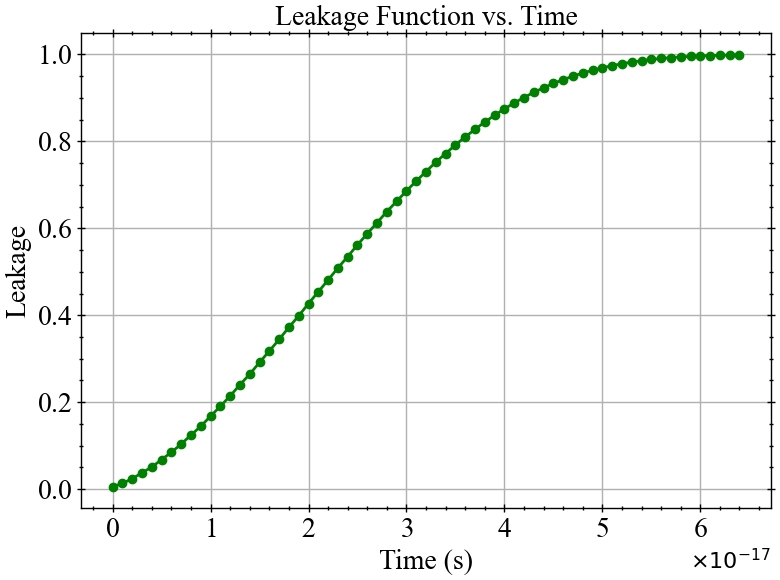}
            \caption{\small  $d=10^{-11}$ m}
        \end{subfigure} &
        \begin{subfigure}[b]{0.3\textwidth}
            \includegraphics[width=\textwidth]{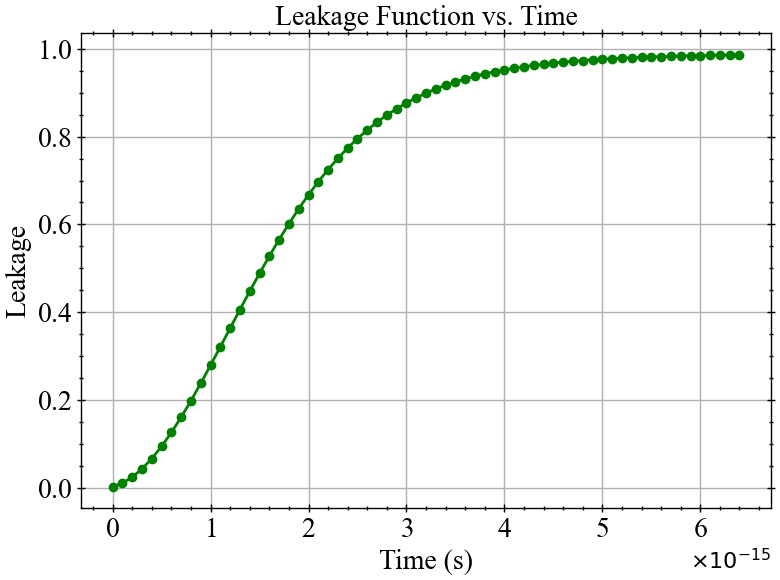}
            \caption{\small  $d=10^{-10}$ m}
        \end{subfigure} \\
         \begin{subfigure}[b]{0.3\textwidth}
            \includegraphics[width=\textwidth]{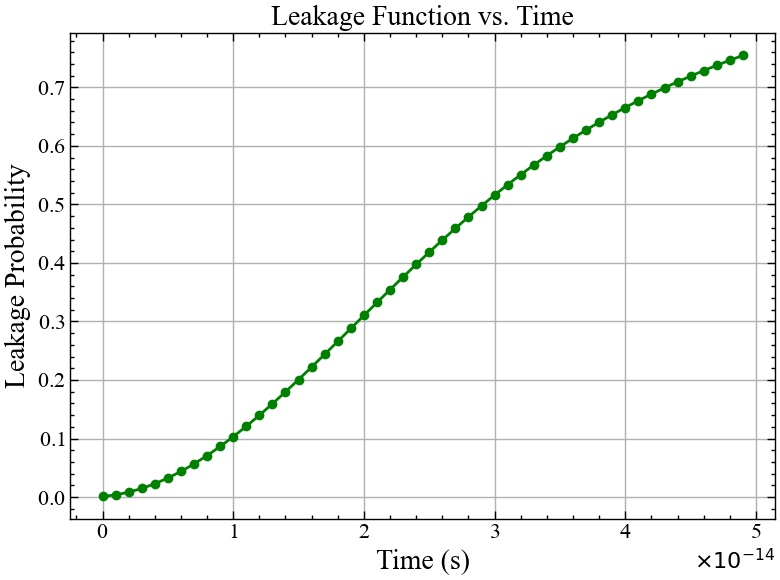}
            \caption{\small $d=10^{-9}$ m}
        \end{subfigure} &
        \begin{subfigure}[b]{0.3\textwidth}
            \includegraphics[width=\textwidth]{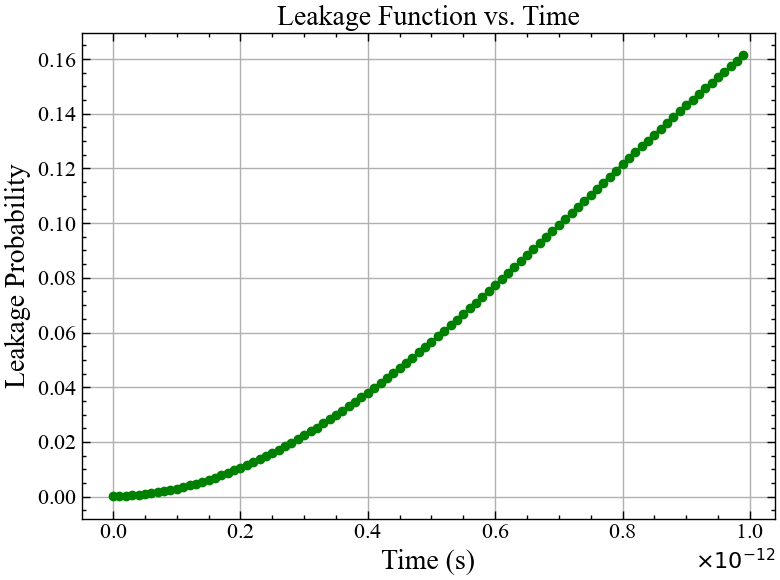}
            \caption{\small  $d=10^{-8}$ m}
        \end{subfigure} &
        \begin{subfigure}[b]{0.3\textwidth}
            \includegraphics[width=\textwidth]{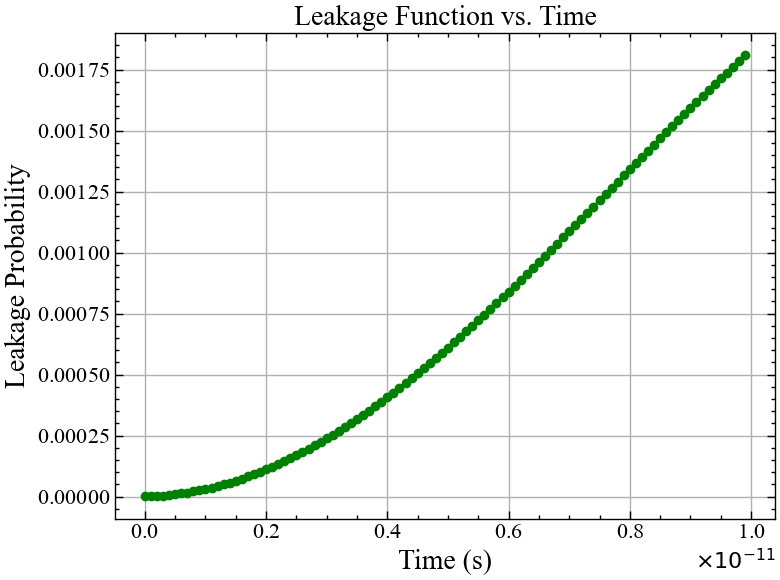}
            \caption{\small  $d=10^{-7}$ m}
            \label{1e-7}
        \end{subfigure} \\
    \end{tabular}
    \caption{Shown are the leakage probability for various proton confinement sizes, from $d=10^{-12}$ m, to $d=10^{-7}$ m, as a function of time.}
    \label{fig4}
\end{figure}

\begin{figure}
    \centering
    \includegraphics[width=0.7\linewidth]{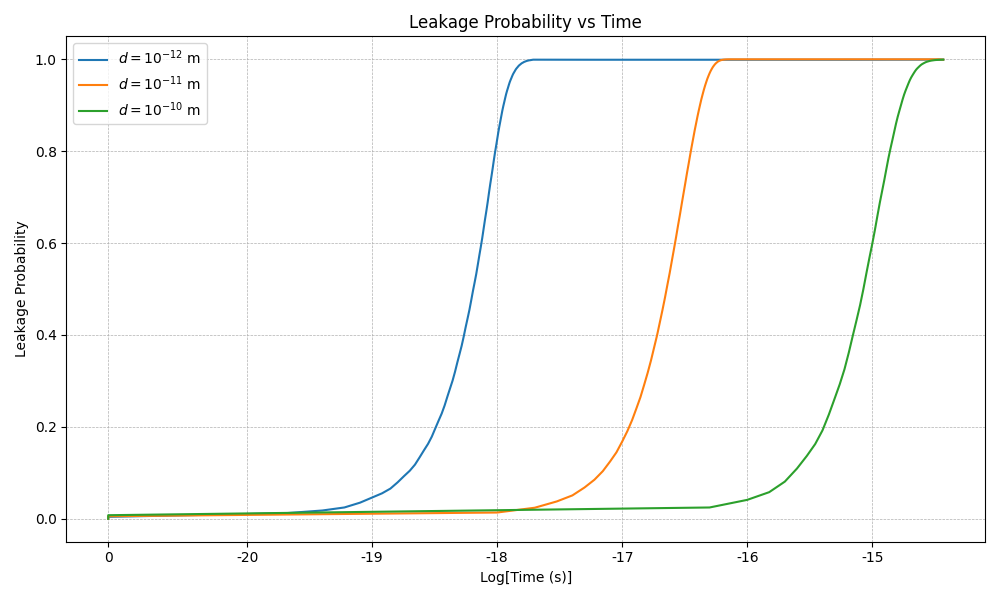}
    \caption{The leakage probability vs. time is shown for different confinement sizes $d$.}
    \label{fig5}
\end{figure}

   In Figure \ref{fig5}, the leakage probability is compared for different confinement sizes on a logarithmic scale. The figure clearly shows that for a given elapsed time, the leakage probability is higher for a smaller confinement size.

\begin{figure}
    \centering
    \includegraphics[width=0.7\linewidth]{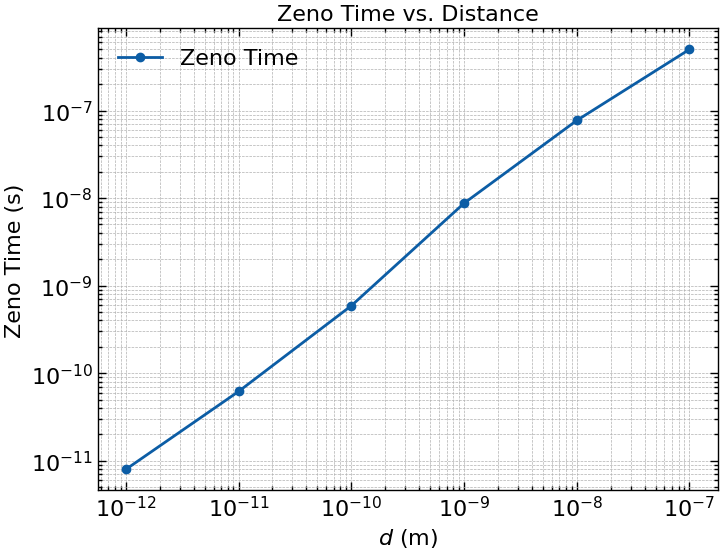}
    \caption{The Zeno time is shown for different confinement sizes $d$. Larger distance $d$ results in longer Zeno time. }
    \label{fig6}
\end{figure}
Finally, we calculate the Zeno time for different values of the confinement sizes for the ground state wavefunction as defined in Eq. (\ref{zt}). The Zeno time is interpreted as how fast the wavefunction decays from its initial state.  In Figure \ref{fig6}, the Zeno time  is shown as a function of the confinement size $d$. For smaller values of $d$, the Zeno time is smaller and for larger values of $d$, the Zeno time is larger. The reason is that the Coulomb repulsion is stronger for smaller $d$ which causes the 2-dimensional wavefunction to spread faster.

Figures \ref{fig4} and \ref{fig5} provide more information. The time $\tau_Y$ during which the two-proton system remains in the confinement region with $Y$ confidence level (CL), is extracted from Figure \ref{fig4}. This gives us an estimation of the measurement frequency one has to consider to confine the ions for a long enough time within the desired spatial region. If we perform $\mathcal{N}$ measurements every $\tau_Y$ time, the probability of finding the system within the confinement region after $t=\mathcal{N} \tau_Y$ is given by $P = Y^\mathcal{N}$. Let us take $Y=95\%$. For $d=10^{-12}$ m, a confinement size where one might expect the probability of the ions fusing to not be negligible, from Figure \ref{fig4} we see that $\tau_Y\sim 10^{-19}$ s.  Therefore, after $100$ measurements the survival probability is $P=(95\%)^{100}\sim 0.5\%$. From Figure \ref{fig6}, the Zeno time for $d=10^{-12}$ m is $\tau_Z\sim 10^{-12}$ s. We observe that even choosing the measurement frequency as small as $\tau_Y\sim 10^{-19} \text{s}< \tau_Z=10^{-12}$ s, does not help remain the protons confined within the confinement region, and the two-ion system after only $10^{-17}$ s decays almost entirely into the exterior of the region. This is to be expected given that we are confining two ions at subatomic distances. Therefore, to quantify the parameter $\tau_Y$ in the condition $\tau_Y=t/\mathcal{N} \ll \tau_Z$ in Eq. (\ref{ztcon}), in order to get successful ion trapping, one should take into account the connection between $\tau_Y$ and the leakage probability from Figures \ref{fig4} and Figures \ref{fig5}.
Let us choose $d=10^{-7}$ m, which is still small, but possibly closer to the type of distances that might be relevant to quantum computing. The corresponding Zeno time as shown in Figure \ref{fig6} is $\tau_Z=5\times 10^{-7}$. As seen in Figure \ref{fig4} (f), if the measurement frequency is chosen to be $\tau_Y\sim 10^{-11}\ll \tau_Z$, the survival probability will be $P\sim99.9\%$. After $100$ measurement, the survival probability becomes $P\sim (99.9\%)^{100}\sim 82\%$ which demonstrates the QZD for the two-ion system. 

In real-world implementations, ideal projective measurements are unattainable. Factors such as finite measurement precision, detector response time, and environmental decoherence place practical limits on how frequently and accurately position measurements can be performed. Furthermore, coupling to the environment can introduce decoherence that may compete with or obscure Zeno effects. These considerations imply that while our simulations demonstrate the ideal limit of Zeno confinement, experimental realizations would require balancing measurement rate, precision, and environmental control.

\section{Conclusion}\label{conc}
The Quantum Zeno Dynamics (QZD) is a generalization of Quantum Zeno Effect (QZE). It states that a quantum system may be frozen in a subset of its Hilbert space, if frequent measurements are performed on the system. If the Hilbert space is taken to be the position of particles, then QZD would help confine them in a desired region of interest. Possible applications range from quantum computing where the trapped ions would form the qubit, to fusion where spatially confining ions to very small distances in a controlled manner is critical. 

An analytical proof of QZD for a single particle is already provided \cite{Facchi_2008}. However, for two or more interacting particles, the calculations become involved and studying the behavior of the systems' wavefunction when measurements are performed is computationally intensive as it cannot be done analytically. In this work, we have performed numerical analysis to demonstrate the QZD using a two-ion system, which can be generalized to multi-ion systems. We have developed Python code which computes the solution to the Schrodinger eigenvalue problem for two-interacting particles confined to a one-dimensional space of length $d$. The Python also evolves the two-particle system in time using the Crank Nicolson method, as well calculating the Zeno time for the quantum system. Furthermore, we define the `leakage function', a metric quantifying the degree to which the wavefunction has spread outside of the region defined by $d$ after a given time. 

As an application of the code, we analyzed the quantum two-proton system and obtained the eigenstate solutions. The ground state wavefunction is depicted in Figure \ref{fig1}. Subsequently, we used the ground state solution of the system as the initial state, and allowed the system to evolve over time, assuming that two ions were initially confined in regions ranging from $d=10^{-12}$ to $d=10^{-7}$ m. Two key parameters, the leakage probability and the Zeno time, were defined and evaluated for different confinement regions with the results presented in Figures \ref{fig4} and \ref{fig5}. Using these results, we could estimate the appropriate measurement frequencies for two specific cases $d=10^{-12}$ and $d=10^{-7}$, in order to keep the two ions trapped within the desired confinement region; hence demonstrating the QZD for the two-ion system.

In summary, the main contribution of this work is the demonstration---through explicit time evolution, Zeno time and leakage analysis---that spatial confinement via QZD is achievable in a realistic two-particle setting with repulsive interaction. This is achieved through a full real-space simulation of two interacting particles subject to repeated spatial projections, revealing how the QZD can suppress Coulomb-driven separation at small scales. In addition, we provide open-source code which can be extended to three-dimensional or many-body systems. Our method bridges analytical theory and computational modeling, offering a platform for exploring measurement-induced confinement in more complex quantum scenarios.

\section*{Acknowledgement}
V.A is grateful to Luke Hopkins and Philip Beadling for their help and advice on Python code creation and debugging. V.A would like to thank Bernard Kay for his constructive feedback on the paper.
P.G. would like to acknowledge support from the International Centre for Theoretical Physics (ICTP) through the Associates Programme (2024-2029). He would also like to thank the TH department at CERN for the hospitality and support where part of this work was done. 

\section*{Authors' Contributions}
V.A. conceptualized the study, developed the computational code, and prepared
the first draft of the manuscript and results. P.G. introduced and implemented
the concept of Zeno time, performed the analysis, and contributed to calculations
throughout the paper. Both authors contributed to writing and editing the
manuscript.

\section*{Data Availability}
The Python code that supports the findings of this study has been deposited in: \url{https://github.com/varqa-abyaneh/Papers/tree/main/Quantum_Zeno_Dynamics_of_Two_Interacting_Particles}

\appendix

\section{Python Code: {\tt 2IonQZD}}\label{code}
In this section we elaborate the Python code provided in appendix \ref{python}, which enables us to study the time-evolution of wavefunction of the two-ion interacting system by utilizing the Crank-Nicolson method. The main code {\tt main.py} uses  a support module called {\tt quantum\_functions.py} in which the electrostatic interaction potential, the essential function to construct the Hamiltonian matrix, and the function to be used in Crank-Nicolson approach, are defined. The necessary system requirements to run the code are the libraries {\tt NumPy}, {\tt SciPy} and {\tt Matplotlib}.

\subsection{\tt quantum\_functions.py}
The module {\tt quantum\_functions.py} defines various functions and parameters that are called and utilized in {\tt main.py}.

\subsection{\tt main.py}
It is assumed that both ions are protons with the masses defined in the code by  {\tt m\_1=1.67262192 e-27} kg and {\tt m\_2=1.67262192 e-27} kg. The electric charge {\tt q\_1=1.60217663 e-19} C and {\tt q\_2=1.60217663 e-19} C. The reduced Planck constant in the code is denoted by {\tt hbar = 1.05457182e-34}. Other parameters in the code are defined as follows: 
\begin{itemize}
\item {\tt epsilon = 1e-15}

This is the regularisation $\epsilon$, defined in Eq. (\ref{eq:Regularisation}) to prevent the potential singularity at $x_1=x_2$.

\item {\tt d = 1.00e-12}
The desired position subspace in units of meter, in which the ions will be confined due to QZD through frequent measurements. Changing this value affects the Zeno time. 

\item {\tt N = 100}

$N$ defines the number of PDE spatial grid points in $x$ and $y$ dimension. The larger values for $N$ leads to more accurate calculations. 

\item {\tt delta\_X = d / N}

This is the spatial grid length in units of meter. 

\item {\tt coulomb\_potential(x, y, q\_1, q\_2, epsilon)}

\end{itemize}

To evaluate the leakage function, the idea is to extend the previous already measured confinement region to {\tt d\_ext} and calculate the probability density that at least one ion is found outside the confinement region {\tt d}. Therefore we define, 

\begin{itemize}
    \item {\tt confinement\_ratio = 2}
    
    The parameter {\tt confinement\_ratio} being set to $2$ here, makes the original confinement distance $d$ double, creating an extended region.

 \item {\tt d\_ext = confinement\_ratio * d}
 
Now, the new extended region is given by {\tt d\_ext}. Setting {\tt confinement ratio = 2} we have {\tt d\_ext=2 d}.

\item {\tt N\_ext = confinement\_ratio * N}

 This is the number of spatial grid points parameter after extending the region from {\tt d}  to {\tt d\_ext} multiplying the original grid points {\tt N} by {\tt confinement\_ratio}. The new grid has the same multiplication factor more points compared to the original grid, as the factor the new region {\tt d\_ext} is larger compared to the original confinement region {\tt d}, keeping to the same grid points resolution before and after the region extension.
 
 \item {\tt delta\_X\_ext = d\_ext / N\_ext}
 
This is the spatial step size for the extended  region {\tt d\_ext}.

 \item {\tt selected\_eigenstate = 0}

 The parameter {\tt selected\_eigenstate} takes values {\tt 0, 1, ...} corresponding to ground state, first exited state, etc, of the Schrodinger equation. 
 Setting {\tt selected\_eigenstate = 0}, the code  will numerically solve the ground state starting off from the lowest-energy eigenstate, as eigenstates are typically indexed starting from 0 in ascending order of energy.

\item  {\tt deltaT = 1e-20}

This is the time step size in the Crank–Nicolson approach for the time evolution of the two-ion quantum system. 

\item {\tt num\_time\_steps = 10}

This is the number of time steps for the Crank–Nicolson method being set to {\tt 10} here, which means that the code will perform  {\tt 10} iterations with time step  {\tt deltaT} for each. 
\end{itemize}

Having introduced the fixed parameters in the code, we now turn to calculation of the ground state wavefunction when 
two ions are initially measured to be in the confined region {\tt d}. To do so, the Hamiltonian $H$ of the quantum system is constructed as a combination of the potential energy (Coulomb interaction) and kinetic energy matrices: {\tt H = T + U}.
\begin{itemize}
\item {\tt V = qf.coulomb\_potential(X, Y, q\_1, q\_2, epsilon)}

This is the two-ion Coulomb potential called from {\tt quantum\_functions.py} imported as {\tt qf} to be used to creat the potential matrix in the next code line. The arguments {\tt X} and {\tt Y} stand for the spatial grid points corresponding to each ion's position, and {\tt q\_1} and {\tt q\_2} are electric charges of the two ions in the system, respectively.
The argument {\tt epsilon} is the regularization parameter defined above. 

\item {\tt U = qf.create\_potential\_matrix(V, N)}

Now, with the {\tt V} defined above we create the potential energy matrix {\tt U} in which it reads the potential {\tt V} and grid size {\tt N} as arguments and returns the potential matrix. 

\item {\tt T = qf.create\_kinetic\_matrix(N, delta\_X, m\_1, m\_2) }

This calls the function {\tt create\_kinetic\_matrix(N, delta\_X, m\_1, m\_2)} as {\tt T} from {\tt quantum\_functions.py} which requires the number of grid {\tt N},  the spatial step size {\tt delta\_X} and the masses of two ions {\tt m\_1} and {\tt m\_2}. As discussed before, the kinetic matrix is constructed based on the finite difference method.
\end{itemize}

In the library {\tt scipy.sparse.linalg}, we make use of the function {\tt eigsh}, which is a matrix eigenvalue solver.  It computes a limited number of eigenvalues and eigenstates of a given sparse Hamiltonian matrix. 

\begin{itemize}
    \item {\tt eigsh(H, k=selected\_eigenstate + 1, which="SM")}
    
    This calculates {\tt k} eigenvalue and eigenstates of the Hamiltonian matrix {\tt H}. By the argument {\tt which="SM"}, the function {\tt eigsh} finds the s"mallest magnitudes "for eigenvalues of {\tt H}. We set {\tt k=selected\_eigenstate+1} i.e. {\tt k=0} to ensure that the ground state is included. The result is two arrays. One is the lowest {\tt k} eigenvaliues and the other is an array of columns each corresponding to an eigenvalue. We choose the eigen state of interest i.e. the ground state, by {\tt eigenvector = eigenvectors[:, selected\_eignestate]}. Note that the ground eigenstate here is stored in a one-dimensional array. 
    
    The resulting grould state eigenvector should be normalized using the line 
    {\tt norm = np.linalg.norm(eigenvector)} which utilizes the function {\tt np.linalg.norm} from {\tt NumPy} library.  
    
    \item {\tt eigenvector\_2D = qf.eigenvector\_1D\_to\_2D(eigenvector, N)}

    This function is defined in {\tt quantum\_functions.py} (imported as {\tt qf}) and is called here in order to make the eigenstate one-dimensional array into a $N\times N$ matrix. From each point on the grid of the wavefunction we extract the probability density of the two ions at the corresponding grid point positions. 
\end{itemize}

Next, we consider the wavefunction evaluated above in the extended grid.  This is to ultimately calculate the leakage function because of the diffusion of the original wavefunction outside the desired confinement region. First, we define a $N_\text{ext} \times N_\text{ext}$ matrix with zero entries by {\tt initial\_state\_2D = np.zeros((N\_ext, N\_ext))}. To put the original confinement grid in the center of the new extended region, or equivalently to set the original wavefunction in the center of the new extended grid, we define the parameter {\tt offset = (N\_ext - N) // 2}. Then,

\begin{itemize}
    \item {\tt initial\_state\_2D[offset : offset + N, offset : offset + N]\\
    = eigenvector\_2D}

This accommodates the original wavefunction  ({\tt eigenvector\_2D}) of $N\times N$ subregion into the center of the extended $N_\text{ext} \times N_\text{ext}$ region, with all surroundings being zero. This padding allows us to observe the wavefunction leakage outside the initial boundaries, by zero entries outside the original grid turning to non-zero values as the sysyem evolves in time. 
\end{itemize}
It is sometimes easier to work with 1D array instead of 2D matrix, so we employ the function {\tt flatten()} to turn the initial eigenstate matrix in the extended region into a 1D array by  {\tt initial\_state\_1D = initial\_state\_2D.flatten()}. This array is normalized then by {\tt linalg.norm}.

In the next step, we begin to evaluate the time evolution using the Crank-Nicolson method. 
First, we need to define the Hamiltonian in the new extended region by giving the potential and kinetic energy matrices in the extended grid. This is done by these lines 
\begin{itemize}
    \item {\tt U\_ext = qf.create\_potential\_matrix(V\_ext, N\_ext)} 
    \item {\tt T\_ext = qf.create\_kinetic\_matrix(N\_ext, delta\_X\_ext, m\_1, m\_2)}
\end{itemize}
where the elecrostatic potential in the extended region is defined by 
\begin{itemize}
    \item {\tt V\_ext = qf.coulomb\_potential(X\_ext, Y\_ext, q\_1, q\_2, epsilon)}
\end{itemize}
To proceed with the time evolution of the wavefunction in the  extended region we need to solve the Eq. (\ref{C-N}) for each grid point. Let us define $A=\left( {I}-iH \Delta t/2\hbar \right)$ and $B=\left( {I}+i H \Delta t/2\hbar  \right)$ as 
\begin{itemize}
    \item {\tt A = sparse.eye(N\_ext**2) - (1j * deltaT / (2 * hbar)) * H\_ext }
\item {\tt B = sparse.eye(N\_ext**2) + (1j * deltaT / (2 * hbar)) * H\_ext}

The parameter time step size {\tt deltaT}, is already set above. The function {\tt eye(N\_ext**2)} is the $N_\text{ext} \times N_\text{ext}$ identity matrix, and {\tt 1j} is the imaginary unit value $i$. 

    \item {\tt psi\_1D\_t, psi\_2D\_t = \\ qf.solve\_future\_states(num\_time\_steps, initial\_state\_1D, A, B, N\_ext)}

    This line calls the function {\tt solve\_future\_states} from {\tt quantum\_functions.py}. This function started from the  computes the evolution of the wavefunction every {\tt DeltaT} for the number of times fixed already by {\tt num\_time\_steps}. The result will be stored in two arrays as 1D and 2D representation of the wavefunction. 

    \item {\tt leakage = qf.calculate\_leakage(psi\_2D\_t[-1], N, N\_ext)}

    This line calls the function {\tt calculate\_leakage} from 
    {\tt quantum\_functions.py}. It calculates the leakage function. That is the probability that at least one of the ions is found outside the initial confinement region. 
\end{itemize}

\section{{\tt 2IonQZD} Code}\label{python}

\begin{lstlisting}[language=Python]
import marshal
import matplotlib.pyplot as plt
import numpy as np
import scienceplots
import structlog
from matplotlib import animation
from matplotlib.animation import FuncAnimation, PillowWriter
from matplotlib.ticker import FuncFormatter
from mpl_toolkits.mplot3d import Axes3D
from scipy import sparse
from scipy.sparse.linalg import eigs, eigsh
import quantum_functions as qf

plt.style.use(["science", "notebook"])

logger = structlog.get_logger()

hbar = 1.05457182e-34
m_1 = 1.67262192e-27  
m_2 = 1.67262192e-27  
q_1 = 1.60217663e-19  
q_2 = 1.60217663e-19  
epsilon = 1e-15  
d = 1.00e-12  
N = 100 
delta_X = d / N 
confinement_ratio = 2 
d_ext = confinement_ratio * d
N_ext = confinement_ratio * N
delta_X_ext = d_ext / N_ext

X, Y = np.mgrid[0 : d : N * 1j, 0 : d : N * 1j]
X_ext, Y_ext = np.mgrid[0 : d_ext : N_ext * 1j, 0 : d_ext : N_ext * 1j]

selected_eignestate = 0
num_time_steps = 2  
deltaT = 1e-20

V = qf.coulomb_potential(X, Y, q_1, q_2, epsilon) 
U = qf.create_potential_matrix(V, N)
T = qf.create_kinetic_matrix(N, delta_X, m_1, m_2)
H = T + U

eigenvalues, eigenvectors = eigsh(H, k=selected_eignestate + 1, which="SM")
print("Energy eigenvalue:", eigenvalues[selected_eignestate])
eigenvector = eigenvectors[:, selected_eignestate]

norm = np.linalg.norm(eigenvector)
print("Norm of the eigenvector:", norm)

eigenvector_2D = qf.eigenvector_1D_to_2D(eigenvector, N)

initial_state_2D = np.zeros((N_ext, N_ext))
offset = (N_ext - N) // 2
initial_state_2D[offset : offset + N, offset : offset + N] = eigenvector_2D

initial_state_1D = initial_state_2D.flatten()

norm = np.linalg.norm(initial_state_1D)
print("Norm of the eigenvector:", norm)

V_ext = qf.coulomb_potential(X_ext, Y_ext, q_1, q_2, epsilon) 
U_ext = qf.create_potential_matrix(V_ext, N_ext)
T_ext = qf.create_kinetic_matrix(N_ext, delta_X_ext, m_1, m_2)
H_ext = T_ext + U_ext

A = sparse.eye(N_ext**2) - (1j * deltaT / (2 * hbar)) * H_ext
B = sparse.eye(N_ext**2) + (1j * deltaT / (2 * hbar)) * H_ext

psi_1D_t, psi_2D_t = qf.solve_future_states(
    num_time_steps, initial_state_1D, A, B, N_ext
)

leakage = qf.calculate_leakage(psi_2D_t[-1], N, N_ext)

qf.graphic_manual_2D_evolve(num_time_steps, psi_2D_t, X_ext, Y_ext, deltaT)

formatter = FuncFormatter(qf.format_ticks)
plt.figure(figsize=(8, 8))
c = plt.pcolormesh(X, Y, eigenvector_2D**2, cmap="nipy_spectral")
plt.xlabel("$x_1$ - position of ion 1 (m)")
plt.ylabel("$x_2$ - position of ion 2 (m)")
plt.gca().xaxis.set_major_formatter(formatter)
plt.gca().yaxis.set_major_formatter(formatter)
plt.gca().tick_params(axis='x', which='major', pad=10)
plt.gca().tick_params(axis='y', which='major', pad=10)
plt.title(r"Two Protons QZE Trapped in a $10^{%d}$m Region" % (np.log10(d)), pad=20, fontsize=20)
plt.colorbar(c, label='probability density')
plt.tight_layout()
plt.show()
X_flat = X.flatten()
Y_flat = Y.flatten()
eigenvector_2D_flat = (eigenvector_2D**2).flatten()
combined_array = np.column_stack((X_flat, Y_flat, eigenvector_2D_flat))
save_file = "2d_wavefunction_data.csv"
np.savetxt(save_file, combined_array, delimiter=",", header="Particle 1 Position,Particle 2 Position,Eigenvector Squared", comments='')
print(f"Data saved to: {save_file}")
\end{lstlisting}

\bibliography{ref.bib}
\bibliographystyle{unsrt}
\end{document}